\documentclass[aps,prl,reprint,superscriptaddress,showpacs]{revtex4-1}
\usepackage{braket}
\usepackage{graphicx}
\usepackage{dcolumn}
\usepackage{bm}
\usepackage{epsfig}
\usepackage{amsmath}
\usepackage[usenames,dvipsnames]{color}
\usepackage{booktabs}
\usepackage{hyperref}
\begin{document}

\title{Quantum oscillations and a non-trivial Berry phase in the noncentrosymmetric topological superconductor candidate BiPd}

\author{Mojammel A. Khan}

\email{mkhan19@lsu.edu}
\affiliation{Department of Physics and Astronomy, Louisiana State University, Baton Rouge, LA 70808}

\author{D. E. Graf}
\affiliation{National High Magnetic Field Laboratory, Tallahassee, FL 32310}

\author{D. Browne}
\affiliation{Department of Physics and Astronomy, Louisiana State University, Baton Rouge, LA 70808}

\author{I. Vekhter}
\affiliation{Department of Physics and Astronomy, Louisiana State University, Baton Rouge, LA 70808}

\author{J. F. DiTusa}
\affiliation{Department of Physics and Astronomy, Louisiana State University, Baton Rouge, LA 70808}

\author{W. Adam Phelan}
\thanks{Current address: Department of Chemistry, The Johns Hopkins University}
\affiliation{Department of Physics and Astronomy, Louisiana State University, Baton Rouge, LA 70808}

\author{D. P. Young}
\affiliation{Department of Physics and Astronomy, Louisiana State University, Baton Rouge, LA 70808}

\date{\today}
\begin{abstract}We report the measurements of de Haas-van Alphen (dHvA) oscillations in the noncentrosymmetric superconductor BiPd. Several pieces of a complex multi-sheet Fermi surface are identified, including a small  pocket (frequency 40 T) which is three dimensional and anisotropic. From the temperature dependence of the amplitude of the oscillations, the cyclotron effective mass is ($0.18$ $\pm$ 0.1) $m_e$. Further analysis showed a non-trivial $\pi$-Berry phase is associated with the 40 T pocket, which strongly supports the presence of topological states in bulk BiPd and may result in topological superconductivity due to the proximity coupling to other bands.


\end{abstract}


\maketitle


Topological materials, such as topological insulators (TI) and superconductors (TSC's), are of great current interest because their complex electronic structure underlies properties that hold  promise in practical applications, such as spintronics and quantum computation~\citep{TIin3D,3dTIBi2Te3,largegapTIwithSO,ZahidHasanTI,QSHinHgTeQW,RevofTIandTSC,SCinCuBiSe}. The essential ingredient is the spin orbit coupling (SOC) that creates non-trivial band states~\citep{largegapTIwithSO,TIin3D}. In a TSC, it is the wave function of the electron pairs that exhibits topological properties, and among the most promising candidates for topological superconductivity are materials with non-centrosymmetric (NCS) crystal structures. In NCS superconductors, the lack of an inversion center lifts the spin degeneracy and, combined with strong antisymmetric SOC, leads to a complex order parameter with mixed spin singlet and spin triplet pairing components~\citep{book,my.paper.Re6Zr,ncs.review}. This can lead to topologically non-trivial superconducting phases, characterized by a full superconducting gap in the bulk, but supporting a protected zero-energy mode at the vortex core (Majorana fermion), 
as well as gapless edge or surface states~\citep{TSC.Majorana1,TSC.Majorana2,TSC.Majorana3}. These latter states are a consequence of Andreev reflection at the boundaries, and hence are referred to as Andreev bound states (ABS)~\citep{TSC.CuxBi2Se3.ABS}. Thus, the NCS family of superconductors is most `auspicious' for hosting topological superconductivity~\citep{TP.of.NCS.majorana.edge,Tp.surface.flat.band.NCS,SAMOKHIN2015385,Tp.gapless.phase.SC,PhysRevB.95.134507}. Thus far, few topological NCS superconductors have been identified~\citep{nodeless.sc.PbTaSe,TSC.PbTaSe2.1,YPtBi.TSC}.

\begin{figure*}[ht]
\centering
\includegraphics[width=0.96\textwidth]{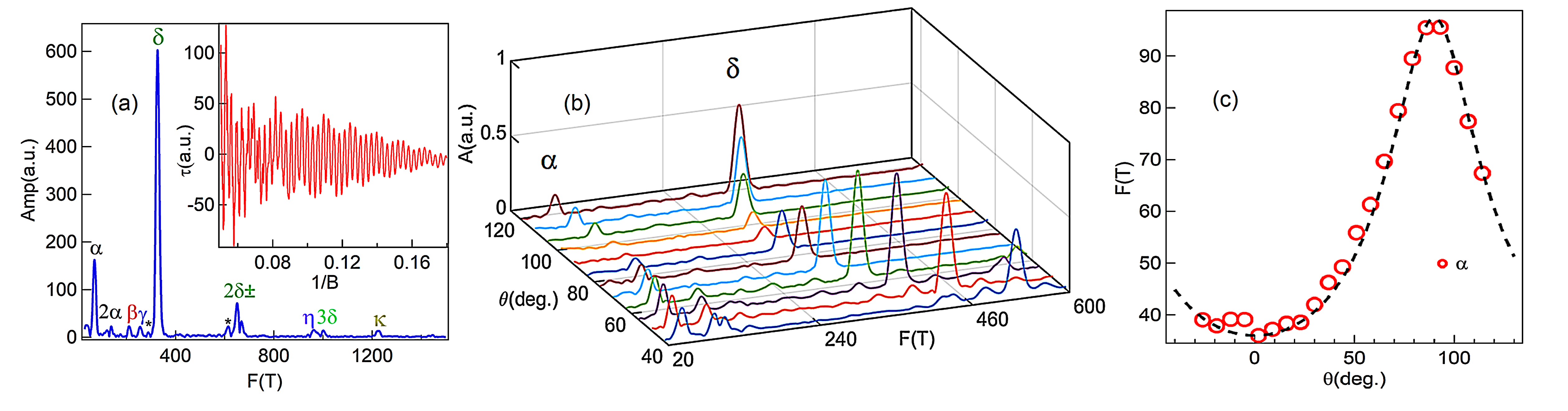}
\caption[Quantum oscillations detected via torque response of the cantilever magnetometer at 0.35 K and corresponding frequencies]{Quantum oscillations detected via torque response, $\tau$, of the cantilever magnetometer at 0.35 K and corresponding frequencies, \textit{F}. (a) Fast Fourier Transform (FFT) of the data shown in the inset for the field range from 5 to 20 T. The primary frequencies $\alpha$, $\beta$, $\gamma$, $\delta$, $\eta$ and $\kappa$, and higher harmonics, are labeled. Inset: Torque response, $\tau$, of the cantilever magnetometer at 0.35 K with field applied  ~70$^\circ$ from the \textit{b}-axis showing dHvA oscillations in inverse field, 1/B. (b) Frequency dispersion in rotating the sample with respect to applied field with angular dependence between $42{^\circ}$ and $119^{\circ}$ with $\theta$ representing the angle between the field and the \textit{b}-axis. A complete list of fundamental frequencies and their dispersion is given in the SM~\citep{supplementary}. (c) Angular dependence of $\alpha$. The frequency can be well described by a 3D anisotropic ellipsoid (dashed line), which  corresponds to a 3D piece of Fermi surface with ellipticity of 2.7.}\label{fft70dgrees}
\end{figure*}


The NCS compound BiPd is a type-II superconductor with a transition temperature of $T_c\sim$3.8 K~\citep{BHATT,benia.BiPd}, which has attracted much interest recently due to the possible unconventional nature of its superconductivity. Here, the unconventional superconducting state is thought to be due to the large SOC caused by the presence of the heavy element Bi~\citep{STM.BiPd}. Surface experiments performed on BiPd thus far show Dirac surface states~\citep{STM.BiPd,benia.BiPd} with anisotropic behavior~\cite{bandhybrid}, albeit these states are far from the chemical potential. A scanning tunneling microscopy (STM) study found a strong zero bias conductance peak (ZBCP)  in the vortex core, and attributed it to a Caroli-de Gennes-Matricon vortex core state~\citep{STM.BiPd,Cd.G.M1}. However, a connection to the Majorana states expected in the vortex cores of topological superconductors could not be established. Directional point contact Andreev reflection (PCAR) measurements also observed a ZBCP, along with evidence for multiple superconducting gaps~\citep{ABS.BiPd}, suggesting an unconventional order parameter. Andreev surface states emerge for certain surface orientations in topologically trivial nodal superconductors, but such an explanation appears to be at odds with thermal
and charge transport measurements, which indicate an isotropic, fully gapped, superconducting state in the bulk~\citep{fullgap.1,fullgap.2}. In fact, many of the properties observed in BiPd are similar to those of other TSC candidates, such as, Cu$_{0.25}$Bi$_{2}$Se$_{3}$~\citep{SCinCuBiSe} and PbTaSe$_{2}$~\citep{nodeless.sc.PbTaSe}. For example, Cu$_{x}$Bi$_{2}$Se$_{3}$ displays a similar ZBCP in PCAR measurements that was attributed to the existence of Majorana fermions and hence topological superconductivity~\citep{TSC.CuxBi2Se3.ABS}. Clearly, further investigations to probe the topological nature of BiPd are desired.

The difficulty in providing the direct `smoking gun' evidence for topology is that few probes directly couple to the phase of the wave function. Recently, quantum oscillations under applied magnetic fields were used to determine the topological nature of the bands in a number of compounds~\citep{landau.fan2,Berry.phase1,Berry.phase2,Berry.phase3,Berry.phase4} through measurements of a non-trivial $\pi$-Berry phase of electronic carriers. To our knowledge no such data have been reported for BiPd, and we fill this gap here.
In this Letter, we report quantum oscillations in single crystalline BiPd measured by torque magnetometry, i.e. the de Haas-van Alpen (dHvA) effect, for different orientations of the magnetic field (\textit{B}). Analysis of the dHvA data suggests a complex three-dimensional Fermi surface composed of multiple sheets. We focus on a pocket with the dHvA frequency of 40 T, where we can track the signal over a wide range of fields and temperatures. We find that these charge carriers have a  very small mass, ($0.18$ $\pm$ 0.01)$m_e$, where $m_e$ is the bare electron mass. We show that a non-trivial Berry phase exists for the same frequency oscillation, indicating Dirac-like properties, including a non-trivial topology. While this particular
pocket of Fermi surface is rather small, we argue that the proximity coupling of
this band with the bands driving superconductivity can lead to a non-trivial
topological nature of the superconducting state.

Single crystals of BiPd were synthesized using a modified Bridgman technique in a radio-frequency induction furnace. Large crystals were produced that were easily cleaved. A cleaved piece was analyzed using X-ray diffraction (XRD) which showed reflections from the (0 $k$ 0) plane only, indicating the cleaved surface was perpendicular to the unique axis ($b$-axis) of its monoclinic structure. A complete XRD analysis, including a temperature dependent investigation showing the transition between the monoclinic and orthorhombic phases of BiPd is given in the Supplementary materials (SM)~\citep{supplementary}.  

A standard four-probe resistivity measurement showed a superconducting transition $T_{c}$ $\sim$ 3.8 K with a residual resistivity ratio (RRR, $\rho_{290K}$/$\rho_{4K}$) greater than 100, indicating excellent crystal quality (see SM~\citep{supplementary} Fig.S2). The measurement of quantum oscillations was performed at the NHMFL in Tallahassee, FL employing a torque magnetometry technique using the 35 T resistive magnet in a cryostat with a base temperature of 350 mK. Details of the measurement techniques are discussed in the SM~\citep{supplementary}.


\begin{figure*}[ht]
\centering
\includegraphics[width=1.0\textwidth]{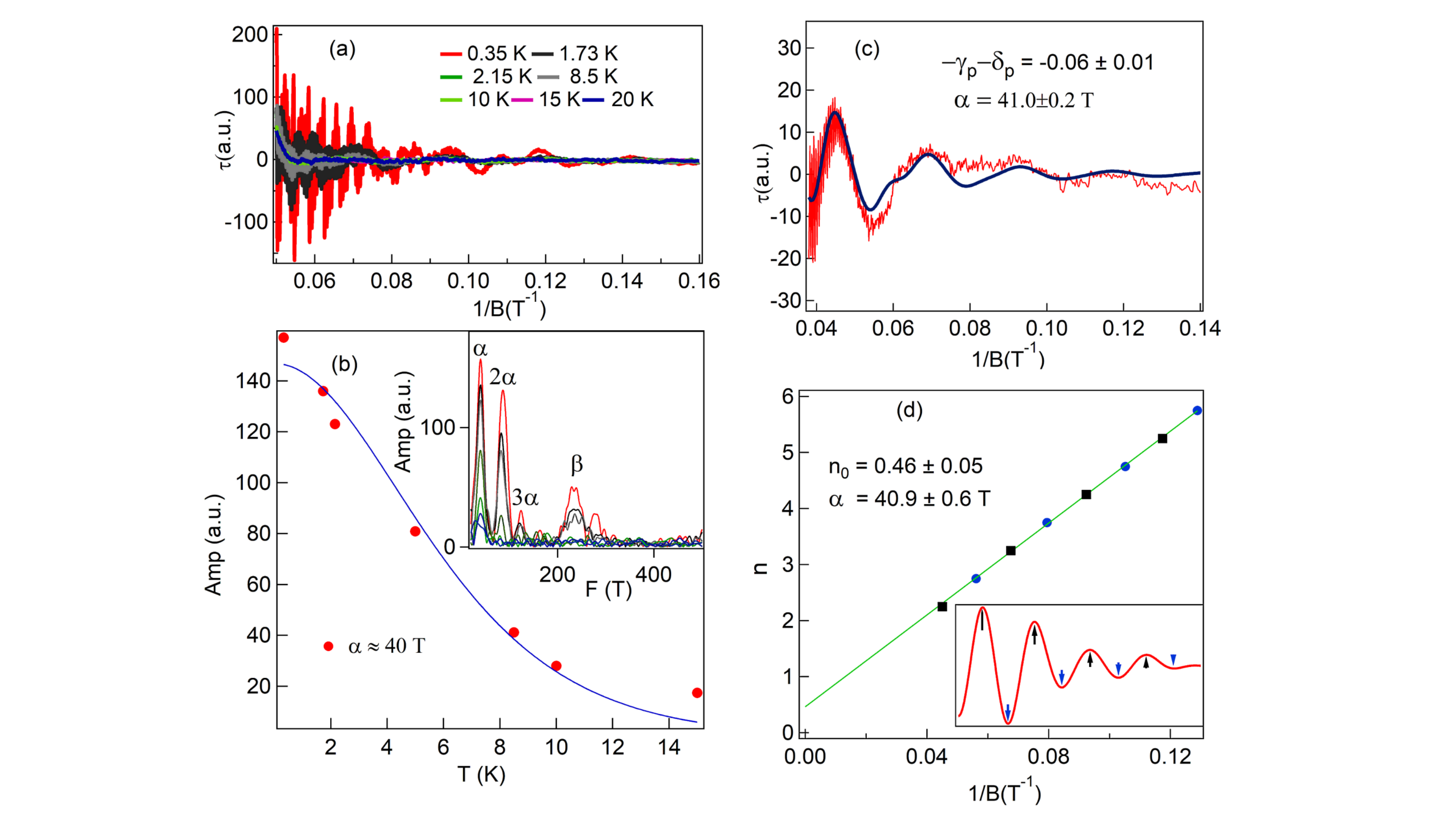}
\caption[Effective mass of the band carriers and Berry phase analysis]{Effective mass of the band carriers and Berry phase analysis. (a) Temperature dependence of the oscillation amplitude, $\tau$, in the field range from 5 to 20 T. (b) Temperature, \textit{T}, dependence of the oscillation amplitude at a field of 8.0 T, along with a fit (solid line) of the LK theory ($R_T$) as described in the text. Inset: FFT of the oscillations shown in (a). (c) The 10 K data (red curve) plotted vs. inverse field. Solid blue line is a fit of the LK theory to the data as described in the text.  Two higher harmonics and the principal frequency were used to fit the data over a broad field range from 7 to 26 T. (d) LL-fan diagram as described in the text. Blue circles represent minima of the oscillations as demonstrated by the arrows in the inset, while black squares represent the maxima. The solid line is a linear fit. Inset: Data after applying a bandpass filter to the torque data around the 40 T frequency. Arrows indicate the positions of the maxima (black) and minima (blue). }\label{mass and phase}.
\end{figure*}

Clear periodic oscillations were observed in the torque measurements, as shown in the inset of Fig.~\ref{fft70dgrees}a, with the magnetic field applied $70^{\circ}$ from the $b$-axis at 0.35 K. At this angular position, the oscillations contain a maximum number of frequencies which were extracted by fast Fourier transformation (FFT). Among those, the 40 T frequency ($\alpha$) is present over the entire angular range (Figs.~\ref{fft70dgrees}b, ~\ref{fft70dgrees}c, and S3). Several other frequencies appear at different field directions and disappear, as the crystal is rotated. 

Rotating the sample with respect to the applied field reveals the dispersion of the dHvA frequencies (Fig.~\ref{fft70dgrees}b). Dispersion for the entire angular range can be found in the SM~\citep{supplementary}. We observed a moderate dispersion in the small frequency $ \alpha $. The value of this frequency increases monotonically to a maximum at $\theta \sim 90^{\circ}$ i.e. field along the $ac$-plane. The angular dependence of the $\alpha$ frequency can be fitted with a form appropriate for a three dimensional  ellipsoid, where the ellipticity was found to be $\sim$ 2.7, as seen in Fig.~\ref{fft70dgrees}c. Thus, $ \alpha $ is an anisotropic, three dimensional (3D) pocket of Fermi surface. 

We analyze the dHvA oscillations by fitting the the oscillatory magnetization to the Lifshitz-Kosevich (LK) formula~\citep{LK1,shoenburg,Berry.phase1}, 

\begin{equation}\label{eq:torque}
\Delta M \propto - B^{\lambda}R_{T}R_{D}R_{S}\, sin\big[ 2\pi(\frac{F}{B}-\gamma_p-\delta_p)\big].
\end{equation}
Here, $R_{T}$ is the thermal damping term, which defines the effective mass of the band carriers, and $R_{D}$ is the Dingle damping factor~\citep{supplementary}. Both of these factors are due to Landau level (LL) broadening, which is caused by the effects of finite temperature, $R_{T}$, and carrier scattering rates, $R_{D}$. The term $R_{S}$ is the spin reduction factor, which is related to the Zeeman effect. A more in depth description of each of these terms is given in the SM~\citep{supplementary}. Finally, the oscillation is described by a sinusoidal term that contains the phase factor (-$\gamma_p$-$\delta_p$), where $\gamma_p$ is related to the Berry phase $\Phi_{B}$ as, $\gamma_p$ = $\frac{1}{2}$ - $\frac{\Phi_{B}}{2\pi}$. The dimension of the Fermi pocket characterizes the value of $\lambda$ and $\delta_p$. For example, for a 3D Fermi surface, the values are 1/2 and $\pm$1/8 (+for minimal and - for a maximal cross section of the constant energy surface), respectively~\citep{LK3}.


The temperature dependence of the oscillation amplitude can be fit by the form for $R_T$ (see SM~\citep{supplementary}) to estimate the carrier effective mass associated with each of the oscillation frequencies. A full temperature dependent data set was collected with the field at an angle of $28^{\circ}$ with respect to the $b$-axis. At this angular position, only the $\alpha$ frequency dominates and persists at temperatures as high as 20 K, as shown in Fig.~\ref{mass and phase}a. This indicates a light mass for the band carriers associated with this frequency. By using the data up to 15 K, we determined the effective mass for this mode as $m^{*}$ = (0.18$\pm$0.01)$m_{e}$ from a fit for $R_T$ described in the SM~\citep{supplementary}, where $m_{e}$ is the mass of a bare electron (Fig.~\ref{mass and phase}b). The amplitudes of the other frequencies drop abruptly with increased temperature as can be seen in Fig.~\ref{mass and phase}a and the inset of Fig.~\ref{mass and phase}b, preventing an accurate estimation of the effective mass associated with the other oscillatory modes. For example, using only a limited number of data points (3 for the frequency mode $\beta$), the effective mass of the frequency was estimated to be (0.55 $\pm$ 0.03)m$_{e}$ (see Fig. S5 in the SM~\citep{supplementary}).

Fig.~\ref{mass and phase}c displays the results of a fit of the LK theory, Eq.~\ref{eq:torque}, of the oscillation amplitude at 10 K. We chose to analyze the data at this particular temperature, since the higher frequencies were substantially damped, so that the oscillations associated with the 40 T frequency were clearly dominant. The best fit of this form to the data results in a frequency of 41.0$\pm$0.2 T, which matches the frequency found in the FFT analysis. In addition, we determine a phase factor, (-$\gamma_p$-$\delta_p$), in Eq.~\ref{eq:torque}, that is small, -0.06$\pm$0.01, indicating a non-trivial Berry phase. If we assume the minimal cross section ($\delta_p = \frac{1}{8}$), we find $\Phi_B = (1.13\pm0.01)\pi$, while for a maximal cross-section ($\delta_p = -\frac{1}{8}$), we find  $\Phi_B = (0.63\pm0.01)\pi$. This result strongly suggests a non-trivial topological structure of the states in the $\alpha$ band.

The value of the Berry phase, can also be extracted from a Landau level (LL) fan diagram, since the density of states is proportional to $dM/dB$~\citep{shoenburg,landau.fan1,landau.fan2}. To further investigate the frequency $ \alpha $, we isolated (inset of Fig.~\ref{mass and phase}d) and indexed the oscillation minima and maxima with integer values \textit{n} and plotted them versus the corresponding inverse field value~\citep{landau.fan1,landau.fan2}. The Berry phase can then be estimated using the intercept $n_{0}$, i.e. $\Phi_{B}$ = 2$\pi$($n_{0}$ $\pm$ $\delta_p$), where $\delta_p$ is defined as above. From the fit to the filtered data (Fig.~\ref{mass and phase}d), the intercept $n_{0}$ is 0.46$\pm$0.05. Using this value, the Berry phase is (1.17$\pm$0.05)$\pi$ for the minimal cross-section or (0.67$\pm$0.05)$\pi$ for the maximal cross-section. These values are consistent with the phase determined from the LK fit. Furthermore, the slope of 40.9$\pm$0.6 T in Fig.~\ref{mass and phase}d is within our measurement uncertainty of the frequency determined from the FFT analysis and the LK fit.

The Dingle temperature $T_{D}$  can be estimated from the magnetic field dependence of the dHvA amplitude  through a comparison to Eq.~\ref{eq:torque}, where the term $R_D$ contains this dependence (see SM~\citep{supplementary}). By fitting the expression for $R_D$ to the field dependence of the $\alpha$ frequency (Fig. S5b), we find $T_{D}$ = 8$\pm$2 K. Using the value of $T_D$, the relaxation time, $\tau_q$, and subsequently the mobility, $\mu_q$, can be estimated through the relations $\tau_{q}$=$\frac{\hbar}{2\pi K_{B}T_{D}}$ and $\mu_{q}$ = $\frac{e\tau_{q}}{m^{*}}$~\citep{shoenburg}. Our data are thus consistent with $\tau_{q}$ = 1.5$\times$ $10^{-13}$ s, and a large mobility $\mu_{q}$ = 1465 cm$^{2}$/Vs. This large mobility is typical of topologically non-trivial states and is found in almost all topological compounds, such as TIs and Dirac materials~\citep{RevofTIandTSC,Rev.TI.Hasan}. Thus, our analysis indicates that the $\alpha$ frequency corresponds to a topologically non-trivial mode with a small effective mass and high mobility. In the quantum oscillation study of the topological superconductor Cu$_{0.25}$Bi$_{2}$Se$_{3}$, the observed Dirac-like band was also found to possess a small effective mass and high mobility~\citep{QO.in.CuBiSe}, and hence the properties of the $\alpha$ band in BiPd are consistent with those of electron bands in other topological superconductors. 
 
It is important to point out that the three-dimensional topological band discovered here is distinct from and not necessarily related to the Dirac surface states observed in BiPd via ARPES and STM studies. The Dirac point of the surface states lies much below the Fermi level i.e. 0.7 eV~\citep{STM.BiPd}, and this large separation in energy scales indicates that the topology of the $\alpha$ mode is independent from that of the surface states.

In summary, we have measured dHvA oscillations in high-quality single crystals of the NCS superconductor BiPd. FFT analysis revealed multiple frequencies associated with a complex Fermi surface. Among those, we focused on the lowest frequency oscillation, $\alpha$, which corresponds to a small pocket of the Fermi surface that we determined to be three-dimensional and anisotropic. Crucially, the analysis of the phase of the oscillations revealed a non-trivial Berry phase associated with the carriers in this pocket. A non trivial Berry phase in the bulk band points toward the possibility of topological states in this compound. 
The small frequency and the concomitant low carrier density in the $\alpha$ band suggest that, if this is the only topological band in BiPd, it can not drive the superconductivity. In that case the topological superconductivity is proximity-driven, and its experimental signatures are sample-dependent, as, for example, strong disorder may suppress the proximity-induced gap. Our data, however, do not preclude the topological nature of other bands in this material, but we can conclusively determine the Berry phase of the $\alpha$ band. The coupling of the non-trivial topology of one or more of the bands to the superconductivity then indicates the realization of topological superconductivity in this compound, similar to the topological superconductor Cu$_{0.25}$Bi$_{2}$Se$_{3}$~\citep{SCinCuBiSe,QO.in.CuBiSe} . Our work shows the first evidence for the topological nature of the carriers in BiPd and provides motivation to further study the nature of the bulk  multiband superconductivity and detailed conditions for the existence of its topological superconducting states.

\acknowledgments
D.P.Y. acknowledges support from the NSF under Grant No. DMR-1306392 for materials synthesis, conductivity and dHvA measurements. J.F.D and D.P.Y. acknowledge the U.S. Department of Energy under EPSCoR Grant No. DE-SC0012432 with additional support from the Louisiana Board of Regents for help with XRD measurements and data analysis. I.V. acknowledges support from NSF Grant No. DMR-1410741 for theoretical analysis. Portions of this work were performed at the NHMFL, which is supported by NSF Cooperative Agreement No. DMR-1157490 and the State of Florida, and  the Analytical Instrumentation Facility (AIF) at North Carolina State University, supported by the State of North Carolina and the NSF award number ECCS-1542015. We acknowledge Dr. C. Loehn at the Shared Instrumentation Facility (SIF), LSU for chemical analysis Dr. C.-C. Chung (AIF) for the temperature dependent X-ray diffraction measurements.

\bibliography{mybib}

\begin{thebibliography}{42}%
\makeatletter
\providecommand \@ifxundefined [1]{%
 \@ifx{#1\undefined}
}%
\providecommand \@ifnum [1]{%
 \ifnum #1\expandafter \@firstoftwo
 \else \expandafter \@secondoftwo
 \fi
}%
\providecommand \@ifx [1]{%
 \ifx #1\expandafter \@firstoftwo
 \else \expandafter \@secondoftwo
 \fi
}%
\providecommand \natexlab [1]{#1}%
\providecommand \enquote  [1]{``#1''}%
\providecommand \bibnamefont  [1]{#1}%
\providecommand \bibfnamefont [1]{#1}%
\providecommand \citenamefont [1]{#1}%
\providecommand \href@noop [0]{\@secondoftwo}%
\providecommand \href [0]{\begingroup \@sanitize@url \@href}%
\providecommand \@href[1]{\@@startlink{#1}\@@href}%
\providecommand \@@href[1]{\endgroup#1\@@endlink}%
\providecommand \@sanitize@url [0]{\catcode `\\12\catcode `\$12\catcode
  `\&12\catcode `\#12\catcode `\^12\catcode `\_12\catcode `\%12\relax}%
\providecommand \@@startlink[1]{}%
\providecommand \@@endlink[0]{}%
\providecommand \url  [0]{\begingroup\@sanitize@url \@url }%
\providecommand \@url [1]{\endgroup\@href {#1}{\urlprefix }}%
\providecommand \urlprefix  [0]{URL }%
\providecommand \Eprint [0]{\href }%
\providecommand \doibase [0]{http://dx.doi.org/}%
\providecommand \selectlanguage [0]{\@gobble}%
\providecommand \bibinfo  [0]{\@secondoftwo}%
\providecommand \bibfield  [0]{\@secondoftwo}%
\providecommand \translation [1]{[#1]}%
\providecommand \BibitemOpen [0]{}%
\providecommand \bibitemStop [0]{}%
\providecommand \bibitemNoStop [0]{.\EOS\space}%
\providecommand \EOS [0]{\spacefactor3000\relax}%
\providecommand \BibitemShut  [1]{\csname bibitem#1\endcsname}%
\let\auto@bib@innerbib\@empty
\bibitem [{\citenamefont {Fu}\ \emph {et~al.}(2007)\citenamefont {Fu},
  \citenamefont {Kane},\ and\ \citenamefont {Mele}}]{TIin3D}%
  \BibitemOpen
  \bibfield  {author} {\bibinfo {author} {\bibfnamefont {L.}~\bibnamefont
  {Fu}}, \bibinfo {author} {\bibfnamefont {C.~L.}\ \bibnamefont {Kane}}, \ and\
  \bibinfo {author} {\bibfnamefont {E.~J.}\ \bibnamefont {Mele}},\ }\href
  {\doibase 10.1103/PhysRevLett.98.106803} {\bibfield  {journal} {\bibinfo
  {journal} {Phys. Rev. Lett.}\ }\textbf {\bibinfo {volume} {98}},\ \bibinfo
  {pages} {106803} (\bibinfo {year} {2007})}\BibitemShut {NoStop}%
\bibitem [{\citenamefont {Chen}\ \emph {et~al.}(2009)\citenamefont {Chen},
  \citenamefont {Analytis}, \citenamefont {Chu}, \citenamefont {Liu},
  \citenamefont {Mo}, \citenamefont {Qi}, \citenamefont {Zhang}, \citenamefont
  {Lu}, \citenamefont {Dai}, \citenamefont {Fang}, \citenamefont {Zhang},
  \citenamefont {Fisher}, \citenamefont {Hussain},\ and\ \citenamefont
  {Shen}}]{3dTIBi2Te3}%
  \BibitemOpen
  \bibfield  {author} {\bibinfo {author} {\bibfnamefont {Y.~L.}\ \bibnamefont
  {Chen}}, \bibinfo {author} {\bibfnamefont {J.~G.}\ \bibnamefont {Analytis}},
  \bibinfo {author} {\bibfnamefont {J.-H.}\ \bibnamefont {Chu}}, \bibinfo
  {author} {\bibfnamefont {Z.~K.}\ \bibnamefont {Liu}}, \bibinfo {author}
  {\bibfnamefont {S.-K.}\ \bibnamefont {Mo}}, \bibinfo {author} {\bibfnamefont
  {X.~L.}\ \bibnamefont {Qi}}, \bibinfo {author} {\bibfnamefont {H.~J.}\
  \bibnamefont {Zhang}}, \bibinfo {author} {\bibfnamefont {D.~H.}\ \bibnamefont
  {Lu}}, \bibinfo {author} {\bibfnamefont {X.}~\bibnamefont {Dai}}, \bibinfo
  {author} {\bibfnamefont {Z.}~\bibnamefont {Fang}}, \bibinfo {author}
  {\bibfnamefont {S.~C.}\ \bibnamefont {Zhang}}, \bibinfo {author}
  {\bibfnamefont {I.~R.}\ \bibnamefont {Fisher}}, \bibinfo {author}
  {\bibfnamefont {Z.}~\bibnamefont {Hussain}}, \ and\ \bibinfo {author}
  {\bibfnamefont {Z.-X.}\ \bibnamefont {Shen}},\ }\href {\doibase
  10.1126/science.1173034} {\bibfield  {journal} {\bibinfo  {journal}
  {Science}\ }\textbf {\bibinfo {volume} {325}},\ \bibinfo {pages} {178}
  (\bibinfo {year} {2009})}\BibitemShut {NoStop}%
\bibitem [{\citenamefont {Xia}\ \emph {et~al.}(2009)\citenamefont {Xia},
  \citenamefont {Qian}, \citenamefont {Hsieh}, \citenamefont {Wray},
  \citenamefont {Pal}, \citenamefont {Lin}, \citenamefont {Bansil},
  \citenamefont {Grauer}, \citenamefont {Hor}, \citenamefont {Cava},\ and\
  \citenamefont {Hasan}}]{largegapTIwithSO}%
  \BibitemOpen
  \bibfield  {author} {\bibinfo {author} {\bibfnamefont {Y.}~\bibnamefont
  {Xia}}, \bibinfo {author} {\bibfnamefont {D.}~\bibnamefont {Qian}}, \bibinfo
  {author} {\bibfnamefont {D.}~\bibnamefont {Hsieh}}, \bibinfo {author}
  {\bibfnamefont {L.}~\bibnamefont {Wray}}, \bibinfo {author} {\bibfnamefont
  {A.}~\bibnamefont {Pal}}, \bibinfo {author} {\bibfnamefont {H.}~\bibnamefont
  {Lin}}, \bibinfo {author} {\bibfnamefont {A.}~\bibnamefont {Bansil}},
  \bibinfo {author} {\bibfnamefont {D.}~\bibnamefont {Grauer}}, \bibinfo
  {author} {\bibfnamefont {Y.~S.}\ \bibnamefont {Hor}}, \bibinfo {author}
  {\bibfnamefont {R.~J.}\ \bibnamefont {Cava}}, \ and\ \bibinfo {author}
  {\bibfnamefont {M.~Z.}\ \bibnamefont {Hasan}},\ }\href
  {http://libezp.lib.lsu.edu/login?url=http://search.ebscohost.com/login.aspx?direct=true&db=a9h&AN=40737212&site=ehost-live&scope=site}
  {\bibfield  {journal} {\bibinfo  {journal} {Nature Physics}\ }\textbf
  {\bibinfo {volume} {5}},\ \bibinfo {pages} {398 } (\bibinfo {year}
  {2009})}\BibitemShut {NoStop}%
\bibitem [{\citenamefont {Hasan}\ and\ \citenamefont
  {Kane}(2010{\natexlab{a}})}]{ZahidHasanTI}%
  \BibitemOpen
  \bibfield  {author} {\bibinfo {author} {\bibfnamefont {M.~Z.}\ \bibnamefont
  {Hasan}}\ and\ \bibinfo {author} {\bibfnamefont {C.~L.}\ \bibnamefont
  {Kane}},\ }\href {\doibase 10.1103/RevModPhys.82.3045} {\bibfield  {journal}
  {\bibinfo  {journal} {Rev. Mod. Phys.}\ }\textbf {\bibinfo {volume} {82}},\
  \bibinfo {pages} {3045} (\bibinfo {year} {2010}{\natexlab{a}})}\BibitemShut
  {NoStop}%
\bibitem [{\citenamefont {K{\"o}nig}\ \emph {et~al.}(2007)\citenamefont
  {K{\"o}nig}, \citenamefont {Wiedmann}, \citenamefont {Br{\"u}ne},
  \citenamefont {Roth}, \citenamefont {Buhmann}, \citenamefont {Molenkamp},
  \citenamefont {Qi},\ and\ \citenamefont {Zhang}}]{QSHinHgTeQW}%
  \BibitemOpen
  \bibfield  {author} {\bibinfo {author} {\bibfnamefont {M.}~\bibnamefont
  {K{\"o}nig}}, \bibinfo {author} {\bibfnamefont {S.}~\bibnamefont {Wiedmann}},
  \bibinfo {author} {\bibfnamefont {C.}~\bibnamefont {Br{\"u}ne}}, \bibinfo
  {author} {\bibfnamefont {A.}~\bibnamefont {Roth}}, \bibinfo {author}
  {\bibfnamefont {H.}~\bibnamefont {Buhmann}}, \bibinfo {author} {\bibfnamefont
  {L.~W.}\ \bibnamefont {Molenkamp}}, \bibinfo {author} {\bibfnamefont {X.-L.}\
  \bibnamefont {Qi}}, \ and\ \bibinfo {author} {\bibfnamefont {S.-C.}\
  \bibnamefont {Zhang}},\ }\href {\doibase 10.1126/science.1148047} {\bibfield
  {journal} {\bibinfo  {journal} {Science}\ }\textbf {\bibinfo {volume}
  {318}},\ \bibinfo {pages} {766} (\bibinfo {year} {2007})}\BibitemShut
  {NoStop}%
\bibitem [{\citenamefont {Qi}\ and\ \citenamefont
  {Zhang}(2011)}]{RevofTIandTSC}%
  \BibitemOpen
  \bibfield  {author} {\bibinfo {author} {\bibfnamefont {X.-L.}\ \bibnamefont
  {Qi}}\ and\ \bibinfo {author} {\bibfnamefont {S.-C.}\ \bibnamefont {Zhang}},\
  }\href {\doibase 10.1103/RevModPhys.83.1057} {\bibfield  {journal} {\bibinfo
  {journal} {Rev. Mod. Phys.}\ }\textbf {\bibinfo {volume} {83}},\ \bibinfo
  {pages} {1057} (\bibinfo {year} {2011})}\BibitemShut {NoStop}%
\bibitem [{\citenamefont {Hor}\ \emph {et~al.}(2010)\citenamefont {Hor},
  \citenamefont {Williams}, \citenamefont {Checkelsky}, \citenamefont
  {Roushan}, \citenamefont {Seo}, \citenamefont {Xu}, \citenamefont
  {Zandbergen}, \citenamefont {Yazdani}, \citenamefont {Ong},\ and\
  \citenamefont {Cava}}]{SCinCuBiSe}%
  \BibitemOpen
  \bibfield  {author} {\bibinfo {author} {\bibfnamefont {Y.~S.}\ \bibnamefont
  {Hor}}, \bibinfo {author} {\bibfnamefont {A.~J.}\ \bibnamefont {Williams}},
  \bibinfo {author} {\bibfnamefont {J.~G.}\ \bibnamefont {Checkelsky}},
  \bibinfo {author} {\bibfnamefont {P.}~\bibnamefont {Roushan}}, \bibinfo
  {author} {\bibfnamefont {J.}~\bibnamefont {Seo}}, \bibinfo {author}
  {\bibfnamefont {Q.}~\bibnamefont {Xu}}, \bibinfo {author} {\bibfnamefont
  {H.~W.}\ \bibnamefont {Zandbergen}}, \bibinfo {author} {\bibfnamefont
  {A.}~\bibnamefont {Yazdani}}, \bibinfo {author} {\bibfnamefont {N.~P.}\
  \bibnamefont {Ong}}, \ and\ \bibinfo {author} {\bibfnamefont {R.~J.}\
  \bibnamefont {Cava}},\ }\href {\doibase 10.1103/PhysRevLett.104.057001}
  {\bibfield  {journal} {\bibinfo  {journal} {Phys. Rev. Lett.}\ }\textbf
  {\bibinfo {volume} {104}},\ \bibinfo {pages} {057001} (\bibinfo {year}
  {2010})}\BibitemShut {NoStop}%
\bibitem [{\citenamefont {Bauer}\ and\ \citenamefont {Sigristz}(2012)}]{book}%
  \BibitemOpen
  \bibfield  {author} {\bibinfo {author} {\bibfnamefont {E.}~\bibnamefont
  {Bauer}}\ and\ \bibinfo {author} {\bibfnamefont {M.}~\bibnamefont
  {Sigristz}},\ }\href@noop {} {\emph {\bibinfo {title} {Non-centrosymmetric
  superconductors: introduction and overview}}}\ (\bibinfo  {publisher}
  {Heidelberg, Springer-Verlag},\ \bibinfo {year} {2012})\BibitemShut {NoStop}%
\bibitem [{\citenamefont {Khan}\ \emph {et~al.}(2016)\citenamefont {Khan},
  \citenamefont {Karki}, \citenamefont {Samanta}, \citenamefont {Browne},
  \citenamefont {Stadler}, \citenamefont {Vekhter}, \citenamefont {Pandey},
  \citenamefont {Adams}, \citenamefont {Young}, \citenamefont {Teknowijoyo},
  \citenamefont {Cho}, \citenamefont {Prozorov},\ and\ \citenamefont
  {Graf}}]{my.paper.Re6Zr}%
  \BibitemOpen
  \bibfield  {author} {\bibinfo {author} {\bibfnamefont {M.~A.}\ \bibnamefont
  {Khan}}, \bibinfo {author} {\bibfnamefont {A.~B.}\ \bibnamefont {Karki}},
  \bibinfo {author} {\bibfnamefont {T.}~\bibnamefont {Samanta}}, \bibinfo
  {author} {\bibfnamefont {D.}~\bibnamefont {Browne}}, \bibinfo {author}
  {\bibfnamefont {S.}~\bibnamefont {Stadler}}, \bibinfo {author} {\bibfnamefont
  {I.}~\bibnamefont {Vekhter}}, \bibinfo {author} {\bibfnamefont
  {A.}~\bibnamefont {Pandey}}, \bibinfo {author} {\bibfnamefont {P.~W.}\
  \bibnamefont {Adams}}, \bibinfo {author} {\bibfnamefont {D.~P.}\ \bibnamefont
  {Young}}, \bibinfo {author} {\bibfnamefont {S.}~\bibnamefont {Teknowijoyo}},
  \bibinfo {author} {\bibfnamefont {K.}~\bibnamefont {Cho}}, \bibinfo {author}
  {\bibfnamefont {R.}~\bibnamefont {Prozorov}}, \ and\ \bibinfo {author}
  {\bibfnamefont {D.~E.}\ \bibnamefont {Graf}},\ }\href {\doibase
  10.1103/PhysRevB.94.144515} {\bibfield  {journal} {\bibinfo  {journal} {Phys.
  Rev. B}\ }\textbf {\bibinfo {volume} {94}},\ \bibinfo {pages} {144515}
  (\bibinfo {year} {2016})}\BibitemShut {NoStop}%
\bibitem [{\citenamefont {Kneidinger}\ \emph {et~al.}(2015)\citenamefont
  {Kneidinger}, \citenamefont {Bauer}, \citenamefont {Zeiringer}, \citenamefont
  {Rogl}, \citenamefont {Blaas-Schenner}, \citenamefont {Reith},\ and\
  \citenamefont {Podloucky}}]{ncs.review}%
  \BibitemOpen
  \bibfield  {author} {\bibinfo {author} {\bibfnamefont {F.}~\bibnamefont
  {Kneidinger}}, \bibinfo {author} {\bibfnamefont {E.}~\bibnamefont {Bauer}},
  \bibinfo {author} {\bibfnamefont {I.}~\bibnamefont {Zeiringer}}, \bibinfo
  {author} {\bibfnamefont {P.}~\bibnamefont {Rogl}}, \bibinfo {author}
  {\bibfnamefont {C.}~\bibnamefont {Blaas-Schenner}}, \bibinfo {author}
  {\bibfnamefont {D.}~\bibnamefont {Reith}}, \ and\ \bibinfo {author}
  {\bibfnamefont {R.}~\bibnamefont {Podloucky}},\ }\href {\doibase
  https://doi.org/10.1016/j.physc.2015.02.016} {\bibfield  {journal} {\bibinfo
  {journal} {Physica C: Superconductivity and its Applications}\ }\textbf
  {\bibinfo {volume} {514}},\ \bibinfo {pages} {388 } (\bibinfo {year}
  {2015})},\ \bibinfo {note} {superconducting Materials: Conventional,
  Unconventional and Undetermined}\BibitemShut {NoStop}%
\bibitem [{\citenamefont {Teo}\ and\ \citenamefont
  {Kane}(2010)}]{TSC.Majorana1}%
  \BibitemOpen
  \bibfield  {author} {\bibinfo {author} {\bibfnamefont {J.~C.~Y.}\
  \bibnamefont {Teo}}\ and\ \bibinfo {author} {\bibfnamefont {C.~L.}\
  \bibnamefont {Kane}},\ }\href {\doibase 10.1103/PhysRevLett.104.046401}
  {\bibfield  {journal} {\bibinfo  {journal} {Phys. Rev. Lett.}\ }\textbf
  {\bibinfo {volume} {104}},\ \bibinfo {pages} {046401} (\bibinfo {year}
  {2010})}\BibitemShut {NoStop}%
\bibitem [{\citenamefont {Qi}\ \emph {et~al.}(2009)\citenamefont {Qi},
  \citenamefont {Hughes}, \citenamefont {Raghu},\ and\ \citenamefont
  {Zhang}}]{TSC.Majorana2}%
  \BibitemOpen
  \bibfield  {author} {\bibinfo {author} {\bibfnamefont {X.-L.}\ \bibnamefont
  {Qi}}, \bibinfo {author} {\bibfnamefont {T.~L.}\ \bibnamefont {Hughes}},
  \bibinfo {author} {\bibfnamefont {S.}~\bibnamefont {Raghu}}, \ and\ \bibinfo
  {author} {\bibfnamefont {S.-C.}\ \bibnamefont {Zhang}},\ }\href {\doibase
  10.1103/PhysRevLett.102.187001} {\bibfield  {journal} {\bibinfo  {journal}
  {Phys. Rev. Lett.}\ }\textbf {\bibinfo {volume} {102}},\ \bibinfo {pages}
  {187001} (\bibinfo {year} {2009})}\BibitemShut {NoStop}%
\bibitem [{\citenamefont {Deng}\ \emph {et~al.}(2012)\citenamefont {Deng},
  \citenamefont {Viola},\ and\ \citenamefont {Ortiz}}]{TSC.Majorana3}%
  \BibitemOpen
  \bibfield  {author} {\bibinfo {author} {\bibfnamefont {S.}~\bibnamefont
  {Deng}}, \bibinfo {author} {\bibfnamefont {L.}~\bibnamefont {Viola}}, \ and\
  \bibinfo {author} {\bibfnamefont {G.}~\bibnamefont {Ortiz}},\ }\href
  {\doibase 10.1103/PhysRevLett.108.036803} {\bibfield  {journal} {\bibinfo
  {journal} {Phys. Rev. Lett.}\ }\textbf {\bibinfo {volume} {108}},\ \bibinfo
  {pages} {036803} (\bibinfo {year} {2012})}\BibitemShut {NoStop}%
\bibitem [{\citenamefont {Sasaki}\ \emph {et~al.}(2011)\citenamefont {Sasaki},
  \citenamefont {Kriener}, \citenamefont {Segawa}, \citenamefont {Yada},
  \citenamefont {Tanaka}, \citenamefont {Sato},\ and\ \citenamefont
  {Ando}}]{TSC.CuxBi2Se3.ABS}%
  \BibitemOpen
  \bibfield  {author} {\bibinfo {author} {\bibfnamefont {S.}~\bibnamefont
  {Sasaki}}, \bibinfo {author} {\bibfnamefont {M.}~\bibnamefont {Kriener}},
  \bibinfo {author} {\bibfnamefont {K.}~\bibnamefont {Segawa}}, \bibinfo
  {author} {\bibfnamefont {K.}~\bibnamefont {Yada}}, \bibinfo {author}
  {\bibfnamefont {Y.}~\bibnamefont {Tanaka}}, \bibinfo {author} {\bibfnamefont
  {M.}~\bibnamefont {Sato}}, \ and\ \bibinfo {author} {\bibfnamefont
  {Y.}~\bibnamefont {Ando}},\ }\href {\doibase 10.1103/PhysRevLett.107.217001}
  {\bibfield  {journal} {\bibinfo  {journal} {Phys. Rev. Lett.}\ }\textbf
  {\bibinfo {volume} {107}},\ \bibinfo {pages} {217001} (\bibinfo {year}
  {2011})}\BibitemShut {NoStop}%
\bibitem [{\citenamefont {Sato}\ and\ \citenamefont
  {Fujimoto}(2009)}]{TP.of.NCS.majorana.edge}%
  \BibitemOpen
  \bibfield  {author} {\bibinfo {author} {\bibfnamefont {M.}~\bibnamefont
  {Sato}}\ and\ \bibinfo {author} {\bibfnamefont {S.}~\bibnamefont
  {Fujimoto}},\ }\href {\doibase 10.1103/PhysRevB.79.094504} {\bibfield
  {journal} {\bibinfo  {journal} {Phys. Rev. B}\ }\textbf {\bibinfo {volume}
  {79}},\ \bibinfo {pages} {094504} (\bibinfo {year} {2009})}\BibitemShut
  {NoStop}%
\bibitem [{\citenamefont {Schnyder}\ and\ \citenamefont
  {Ryu}(2011)}]{Tp.surface.flat.band.NCS}%
  \BibitemOpen
  \bibfield  {author} {\bibinfo {author} {\bibfnamefont {A.~P.}\ \bibnamefont
  {Schnyder}}\ and\ \bibinfo {author} {\bibfnamefont {S.}~\bibnamefont {Ryu}},\
  }\href {\doibase 10.1103/PhysRevB.84.060504} {\bibfield  {journal} {\bibinfo
  {journal} {Phys. Rev. B}\ }\textbf {\bibinfo {volume} {84}},\ \bibinfo
  {pages} {060504} (\bibinfo {year} {2011})}\BibitemShut {NoStop}%
\bibitem [{\citenamefont {Samokhin}(2015)}]{SAMOKHIN2015385}%
  \BibitemOpen
  \bibfield  {author} {\bibinfo {author} {\bibfnamefont {K.}~\bibnamefont
  {Samokhin}},\ }\href {\doibase https://doi.org/10.1016/j.aop.2015.04.024}
  {\bibfield  {journal} {\bibinfo  {journal} {Annals of Physics}\ }\textbf
  {\bibinfo {volume} {359}},\ \bibinfo {pages} {385 } (\bibinfo {year}
  {2015})}\BibitemShut {NoStop}%
\bibitem [{\citenamefont {B\'eri}(2010)}]{Tp.gapless.phase.SC}%
  \BibitemOpen
  \bibfield  {author} {\bibinfo {author} {\bibfnamefont {B.}~\bibnamefont
  {B\'eri}},\ }\href {\doibase 10.1103/PhysRevB.81.134515} {\bibfield
  {journal} {\bibinfo  {journal} {Phys. Rev. B}\ }\textbf {\bibinfo {volume}
  {81}},\ \bibinfo {pages} {134515} (\bibinfo {year} {2010})}\BibitemShut
  {NoStop}%
\bibitem [{\citenamefont {Daido}\ and\ \citenamefont
  {Yanase}(2017)}]{PhysRevB.95.134507}%
  \BibitemOpen
  \bibfield  {author} {\bibinfo {author} {\bibfnamefont {A.}~\bibnamefont
  {Daido}}\ and\ \bibinfo {author} {\bibfnamefont {Y.}~\bibnamefont {Yanase}},\
  }\href {\doibase 10.1103/PhysRevB.95.134507} {\bibfield  {journal} {\bibinfo
  {journal} {Phys. Rev. B}\ }\textbf {\bibinfo {volume} {95}},\ \bibinfo
  {pages} {134507} (\bibinfo {year} {2017})}\BibitemShut {NoStop}%
\bibitem [{\citenamefont {Pang}\ \emph {et~al.}(2016)\citenamefont {Pang},
  \citenamefont {Smidman}, \citenamefont {Zhao}, \citenamefont {Wang},
  \citenamefont {Weng}, \citenamefont {Che}, \citenamefont {Chen},
  \citenamefont {Lu}, \citenamefont {Chen},\ and\ \citenamefont
  {Yuan}}]{nodeless.sc.PbTaSe}%
  \BibitemOpen
  \bibfield  {author} {\bibinfo {author} {\bibfnamefont {G.~M.}\ \bibnamefont
  {Pang}}, \bibinfo {author} {\bibfnamefont {M.}~\bibnamefont {Smidman}},
  \bibinfo {author} {\bibfnamefont {L.~X.}\ \bibnamefont {Zhao}}, \bibinfo
  {author} {\bibfnamefont {Y.~F.}\ \bibnamefont {Wang}}, \bibinfo {author}
  {\bibfnamefont {Z.~F.}\ \bibnamefont {Weng}}, \bibinfo {author}
  {\bibfnamefont {L.~Q.}\ \bibnamefont {Che}}, \bibinfo {author} {\bibfnamefont
  {Y.}~\bibnamefont {Chen}}, \bibinfo {author} {\bibfnamefont {X.}~\bibnamefont
  {Lu}}, \bibinfo {author} {\bibfnamefont {G.~F.}\ \bibnamefont {Chen}}, \ and\
  \bibinfo {author} {\bibfnamefont {H.~Q.}\ \bibnamefont {Yuan}},\ }\href
  {\doibase 10.1103/PhysRevB.93.060506} {\bibfield  {journal} {\bibinfo
  {journal} {Phys. Rev. B}\ }\textbf {\bibinfo {volume} {93}},\ \bibinfo
  {pages} {060506} (\bibinfo {year} {2016})}\BibitemShut {NoStop}%
\bibitem [{\citenamefont {Guan}\ \emph {et~al.}(2016)\citenamefont {Guan},
  \citenamefont {Chen}, \citenamefont {Chu}, \citenamefont {Sankar},
  \citenamefont {Chou}, \citenamefont {Jeng}, \citenamefont {Chang},\ and\
  \citenamefont {Chuang}}]{TSC.PbTaSe2.1}%
  \BibitemOpen
  \bibfield  {author} {\bibinfo {author} {\bibfnamefont {S.-Y.}\ \bibnamefont
  {Guan}}, \bibinfo {author} {\bibfnamefont {P.-J.}\ \bibnamefont {Chen}},
  \bibinfo {author} {\bibfnamefont {M.-W.}\ \bibnamefont {Chu}}, \bibinfo
  {author} {\bibfnamefont {R.}~\bibnamefont {Sankar}}, \bibinfo {author}
  {\bibfnamefont {F.}~\bibnamefont {Chou}}, \bibinfo {author} {\bibfnamefont
  {H.-T.}\ \bibnamefont {Jeng}}, \bibinfo {author} {\bibfnamefont {C.-S.}\
  \bibnamefont {Chang}}, \ and\ \bibinfo {author} {\bibfnamefont {T.-M.}\
  \bibnamefont {Chuang}},\ }\href
  {http://advances.sciencemag.org/content/advances/2/11/e1600894.full.pdf}
  {\bibfield  {journal} {\bibinfo  {journal} {Science Advances}\ }\textbf
  {\bibinfo {volume} {2}},\ \bibinfo {pages} {e1600894} (\bibinfo {year}
  {2016})}\BibitemShut {NoStop}%
\bibitem [{\citenamefont {Liu}\ \emph {et~al.}(2016)\citenamefont {Liu},
  \citenamefont {Yang}, \citenamefont {Wu}, \citenamefont {Shekhar},
  \citenamefont {Jiang}, \citenamefont {Yang}, \citenamefont {Zhang},
  \citenamefont {Mo}, \citenamefont {Hussain}, \citenamefont {Yan} \emph
  {et~al.}}]{YPtBi.TSC}%
  \BibitemOpen
  \bibfield  {author} {\bibinfo {author} {\bibfnamefont {Z.}~\bibnamefont
  {Liu}}, \bibinfo {author} {\bibfnamefont {L.}~\bibnamefont {Yang}}, \bibinfo
  {author} {\bibfnamefont {S.-C.}\ \bibnamefont {Wu}}, \bibinfo {author}
  {\bibfnamefont {C.}~\bibnamefont {Shekhar}}, \bibinfo {author} {\bibfnamefont
  {J.}~\bibnamefont {Jiang}}, \bibinfo {author} {\bibfnamefont
  {H.}~\bibnamefont {Yang}}, \bibinfo {author} {\bibfnamefont {Y.}~\bibnamefont
  {Zhang}}, \bibinfo {author} {\bibfnamefont {S.-K.}\ \bibnamefont {Mo}},
  \bibinfo {author} {\bibfnamefont {Z.}~\bibnamefont {Hussain}}, \bibinfo
  {author} {\bibfnamefont {B.}~\bibnamefont {Yan}},  \emph {et~al.},\ }\href
  {https://www.nature.com/articles/ncomms12924} {\bibfield  {journal} {\bibinfo
   {journal} {Nature communications}\ }\textbf {\bibinfo {volume} {7}}
  (\bibinfo {year} {2016})}\BibitemShut {NoStop}%
\bibitem [{\citenamefont {Supplemetary}()}]{supplementary}%
  \BibitemOpen
  \bibfield  {author} {\bibinfo {author} {\bibnamefont {Supplemetary}},\
  }\href@noop {} {\ }\BibitemShut {NoStop}%
\bibitem [{\citenamefont {Bhatt}\ and\ \citenamefont {Schubert}(1980)}]{BHATT}%
  \BibitemOpen
  \bibfield  {author} {\bibinfo {author} {\bibfnamefont {Y.}~\bibnamefont
  {Bhatt}}\ and\ \bibinfo {author} {\bibfnamefont {K.}~\bibnamefont
  {Schubert}},\ }\href {\doibase
  http://dx.doi.org/10.1016/0022-5088(80)90285-4} {\bibfield  {journal}
  {\bibinfo  {journal} {Journal of the Less Common Metals}\ }\textbf {\bibinfo
  {volume} {70}},\ \bibinfo {pages} {P39 } (\bibinfo {year}
  {1980})}\BibitemShut {NoStop}%
\bibitem [{\citenamefont {Benia}\ \emph {et~al.}(2016)\citenamefont {Benia},
  \citenamefont {Rampi}, \citenamefont {Trainer}, \citenamefont {Yim},
  \citenamefont {Maldonado}, \citenamefont {Peets}, \citenamefont {St\"ohr},
  \citenamefont {Starke}, \citenamefont {Kern}, \citenamefont {Yaresko},
  \citenamefont {Levy}, \citenamefont {Damascelli}, \citenamefont {Ast},
  \citenamefont {Schnyder},\ and\ \citenamefont {Wahl}}]{benia.BiPd}%
  \BibitemOpen
  \bibfield  {author} {\bibinfo {author} {\bibfnamefont {H.~M.}\ \bibnamefont
  {Benia}}, \bibinfo {author} {\bibfnamefont {E.}~\bibnamefont {Rampi}},
  \bibinfo {author} {\bibfnamefont {C.}~\bibnamefont {Trainer}}, \bibinfo
  {author} {\bibfnamefont {C.~M.}\ \bibnamefont {Yim}}, \bibinfo {author}
  {\bibfnamefont {A.}~\bibnamefont {Maldonado}}, \bibinfo {author}
  {\bibfnamefont {D.~C.}\ \bibnamefont {Peets}}, \bibinfo {author}
  {\bibfnamefont {A.}~\bibnamefont {St\"ohr}}, \bibinfo {author} {\bibfnamefont
  {U.}~\bibnamefont {Starke}}, \bibinfo {author} {\bibfnamefont
  {K.}~\bibnamefont {Kern}}, \bibinfo {author} {\bibfnamefont {A.}~\bibnamefont
  {Yaresko}}, \bibinfo {author} {\bibfnamefont {G.}~\bibnamefont {Levy}},
  \bibinfo {author} {\bibfnamefont {A.}~\bibnamefont {Damascelli}}, \bibinfo
  {author} {\bibfnamefont {C.~R.}\ \bibnamefont {Ast}}, \bibinfo {author}
  {\bibfnamefont {A.~P.}\ \bibnamefont {Schnyder}}, \ and\ \bibinfo {author}
  {\bibfnamefont {P.}~\bibnamefont {Wahl}},\ }\href {\doibase
  10.1103/PhysRevB.94.121407} {\bibfield  {journal} {\bibinfo  {journal} {Phys.
  Rev. B}\ }\textbf {\bibinfo {volume} {94}},\ \bibinfo {pages} {121407}
  (\bibinfo {year} {2016})}\BibitemShut {NoStop}%
\bibitem [{\citenamefont {Sun}\ \emph {et~al.}(2015)\citenamefont {Sun},
  \citenamefont {Enayat}, \citenamefont {Maldonado}, \citenamefont {Lithgow},
  \citenamefont {Yelland}, \citenamefont {Peets}, \citenamefont {Yaresko},
  \citenamefont {Schnyder},\ and\ \citenamefont {Wahl}}]{STM.BiPd}%
  \BibitemOpen
  \bibfield  {author} {\bibinfo {author} {\bibfnamefont {Z.}~\bibnamefont
  {Sun}}, \bibinfo {author} {\bibfnamefont {M.}~\bibnamefont {Enayat}},
  \bibinfo {author} {\bibfnamefont {A.}~\bibnamefont {Maldonado}}, \bibinfo
  {author} {\bibfnamefont {C.}~\bibnamefont {Lithgow}}, \bibinfo {author}
  {\bibfnamefont {E.}~\bibnamefont {Yelland}}, \bibinfo {author} {\bibfnamefont
  {D.~C.}\ \bibnamefont {Peets}}, \bibinfo {author} {\bibfnamefont
  {A.}~\bibnamefont {Yaresko}}, \bibinfo {author} {\bibfnamefont {A.~P.}\
  \bibnamefont {Schnyder}}, \ and\ \bibinfo {author} {\bibfnamefont
  {P.}~\bibnamefont {Wahl}},\ }\href {\doibase doi:10.1038/ncomms7633}
  {\bibfield  {journal} {\bibinfo  {journal} {Nat. Commun.}\ }\textbf {\bibinfo
  {volume} {6}},\ \bibinfo {pages} {6633} (\bibinfo {year} {2015})}\BibitemShut
  {NoStop}%
\bibitem [{\citenamefont {Thirupathaiah}\ \emph {et~al.}(2016)\citenamefont
  {Thirupathaiah}, \citenamefont {Ghosh}, \citenamefont {Jha}, \citenamefont
  {Rienks}, \citenamefont {Dolui}, \citenamefont {Ravi~Kishore}, \citenamefont
  {B\"uchner}, \citenamefont {Das}, \citenamefont {Awana}, \citenamefont
  {Sarma},\ and\ \citenamefont {Fink}}]{bandhybrid}%
  \BibitemOpen
  \bibfield  {author} {\bibinfo {author} {\bibfnamefont {S.}~\bibnamefont
  {Thirupathaiah}}, \bibinfo {author} {\bibfnamefont {S.}~\bibnamefont
  {Ghosh}}, \bibinfo {author} {\bibfnamefont {R.}~\bibnamefont {Jha}}, \bibinfo
  {author} {\bibfnamefont {E.~D.~L.}\ \bibnamefont {Rienks}}, \bibinfo {author}
  {\bibfnamefont {K.}~\bibnamefont {Dolui}}, \bibinfo {author} {\bibfnamefont
  {V.~V.}\ \bibnamefont {Ravi~Kishore}}, \bibinfo {author} {\bibfnamefont
  {B.}~\bibnamefont {B\"uchner}}, \bibinfo {author} {\bibfnamefont
  {T.}~\bibnamefont {Das}}, \bibinfo {author} {\bibfnamefont {V.~P.~S.}\
  \bibnamefont {Awana}}, \bibinfo {author} {\bibfnamefont {D.~D.}\ \bibnamefont
  {Sarma}}, \ and\ \bibinfo {author} {\bibfnamefont {J.}~\bibnamefont {Fink}},\
  }\href {\doibase 10.1103/PhysRevLett.117.177001} {\bibfield  {journal}
  {\bibinfo  {journal} {Phys. Rev. Lett.}\ }\textbf {\bibinfo {volume} {117}},\
  \bibinfo {pages} {177001} (\bibinfo {year} {2016})}\BibitemShut {NoStop}%
\bibitem [{\citenamefont {Caroli}\ \emph {et~al.}(1964)\citenamefont {Caroli},
  \citenamefont {Gennes},\ and\ \citenamefont {Matricon}}]{Cd.G.M1}%
  \BibitemOpen
  \bibfield  {author} {\bibinfo {author} {\bibfnamefont {C.}~\bibnamefont
  {Caroli}}, \bibinfo {author} {\bibfnamefont {P.~D.}\ \bibnamefont {Gennes}},
  \ and\ \bibinfo {author} {\bibfnamefont {J.}~\bibnamefont {Matricon}},\
  }\href {\doibase http://doi.org/10.1016/0031-9163(64)90375-0} {\bibfield
  {journal} {\bibinfo  {journal} {Physics Letters}\ }\textbf {\bibinfo {volume}
  {9}},\ \bibinfo {pages} {307 } (\bibinfo {year} {1964})}\BibitemShut
  {NoStop}%
\bibitem [{\citenamefont {Mondal}\ \emph {et~al.}(2012)\citenamefont {Mondal},
  \citenamefont {Joshi}, \citenamefont {Kumar}, \citenamefont {Kamlapure},
  \citenamefont {Ganguli}, \citenamefont {Thamizhavel}, \citenamefont {Mandal},
  \citenamefont {Ramakrishnan},\ and\ \citenamefont {Raychaudhuri}}]{ABS.BiPd}%
  \BibitemOpen
  \bibfield  {author} {\bibinfo {author} {\bibfnamefont {M.}~\bibnamefont
  {Mondal}}, \bibinfo {author} {\bibfnamefont {B.}~\bibnamefont {Joshi}},
  \bibinfo {author} {\bibfnamefont {S.}~\bibnamefont {Kumar}}, \bibinfo
  {author} {\bibfnamefont {A.}~\bibnamefont {Kamlapure}}, \bibinfo {author}
  {\bibfnamefont {S.~C.}\ \bibnamefont {Ganguli}}, \bibinfo {author}
  {\bibfnamefont {A.}~\bibnamefont {Thamizhavel}}, \bibinfo {author}
  {\bibfnamefont {S.~S.}\ \bibnamefont {Mandal}}, \bibinfo {author}
  {\bibfnamefont {S.}~\bibnamefont {Ramakrishnan}}, \ and\ \bibinfo {author}
  {\bibfnamefont {P.}~\bibnamefont {Raychaudhuri}},\ }\href {\doibase
  10.1103/PhysRevB.86.094520} {\bibfield  {journal} {\bibinfo  {journal} {Phys.
  Rev. B}\ }\textbf {\bibinfo {volume} {86}},\ \bibinfo {pages} {094520}
  (\bibinfo {year} {2012})}\BibitemShut {NoStop}%
\bibitem [{\citenamefont {Yan}\ \emph {et~al.}(2016)\citenamefont {Yan},
  \citenamefont {Xu}, \citenamefont {He}, \citenamefont {Dong}, \citenamefont
  {Cho}, \citenamefont {Peets}, \citenamefont {Park},\ and\ \citenamefont
  {Li}}]{fullgap.1}%
  \BibitemOpen
  \bibfield  {author} {\bibinfo {author} {\bibfnamefont {X.~B.}\ \bibnamefont
  {Yan}}, \bibinfo {author} {\bibfnamefont {Y.}~\bibnamefont {Xu}}, \bibinfo
  {author} {\bibfnamefont {L.~P.}\ \bibnamefont {He}}, \bibinfo {author}
  {\bibfnamefont {J.~K.}\ \bibnamefont {Dong}}, \bibinfo {author}
  {\bibfnamefont {H.}~\bibnamefont {Cho}}, \bibinfo {author} {\bibfnamefont
  {D.~C.}\ \bibnamefont {Peets}}, \bibinfo {author} {\bibfnamefont {J.-G.}\
  \bibnamefont {Park}}, \ and\ \bibinfo {author} {\bibfnamefont {S.~Y.}\
  \bibnamefont {Li}},\ }\href {http://stacks.iop.org/0953-2048/29/i=6/a=065001}
  {\bibfield  {journal} {\bibinfo  {journal} {Superconductor Science and
  Technology}\ }\textbf {\bibinfo {volume} {29}},\ \bibinfo {pages} {065001}
  (\bibinfo {year} {2016})}\BibitemShut {NoStop}%
\bibitem [{\citenamefont {Joshi}\ \emph {et~al.}(2011)\citenamefont {Joshi},
  \citenamefont {Thamizhavel},\ and\ \citenamefont {Ramakrishnan}}]{fullgap.2}%
  \BibitemOpen
  \bibfield  {author} {\bibinfo {author} {\bibfnamefont {B.}~\bibnamefont
  {Joshi}}, \bibinfo {author} {\bibfnamefont {A.}~\bibnamefont {Thamizhavel}},
  \ and\ \bibinfo {author} {\bibfnamefont {S.}~\bibnamefont {Ramakrishnan}},\
  }\href {\doibase 10.1103/PhysRevB.84.064518} {\bibfield  {journal} {\bibinfo
  {journal} {Phys. Rev. B}\ }\textbf {\bibinfo {volume} {84}},\ \bibinfo
  {pages} {064518} (\bibinfo {year} {2011})}\BibitemShut {NoStop}%
\bibitem [{\citenamefont {Ando}(2013)}]{landau.fan2}%
  \BibitemOpen
  \bibfield  {author} {\bibinfo {author} {\bibfnamefont {Y.}~\bibnamefont
  {Ando}},\ }\href {\doibase 10.7566/JPSJ.82.102001} {\bibfield  {journal}
  {\bibinfo  {journal} {Journal of the Physical Society of Japan}\ }\textbf
  {\bibinfo {volume} {82}},\ \bibinfo {pages} {102001} (\bibinfo {year}
  {2013})}\BibitemShut {NoStop}%
\bibitem [{\citenamefont {Hu}\ \emph {et~al.}(2016{\natexlab{a}})\citenamefont
  {Hu}, \citenamefont {Tang}, \citenamefont {Liu}, \citenamefont {Liu},
  \citenamefont {Zhu}, \citenamefont {Graf}, \citenamefont {Myhro},
  \citenamefont {Tran}, \citenamefont {Lau}, \citenamefont {Wei},\ and\
  \citenamefont {Mao}}]{Berry.phase1}%
  \BibitemOpen
  \bibfield  {author} {\bibinfo {author} {\bibfnamefont {J.}~\bibnamefont
  {Hu}}, \bibinfo {author} {\bibfnamefont {Z.}~\bibnamefont {Tang}}, \bibinfo
  {author} {\bibfnamefont {J.}~\bibnamefont {Liu}}, \bibinfo {author}
  {\bibfnamefont {X.}~\bibnamefont {Liu}}, \bibinfo {author} {\bibfnamefont
  {Y.}~\bibnamefont {Zhu}}, \bibinfo {author} {\bibfnamefont {D.}~\bibnamefont
  {Graf}}, \bibinfo {author} {\bibfnamefont {K.}~\bibnamefont {Myhro}},
  \bibinfo {author} {\bibfnamefont {S.}~\bibnamefont {Tran}}, \bibinfo {author}
  {\bibfnamefont {C.~N.}\ \bibnamefont {Lau}}, \bibinfo {author} {\bibfnamefont
  {J.}~\bibnamefont {Wei}}, \ and\ \bibinfo {author} {\bibfnamefont
  {Z.}~\bibnamefont {Mao}},\ }\href {\doibase 10.1103/PhysRevLett.117.016602}
  {\bibfield  {journal} {\bibinfo  {journal} {Phys. Rev. Lett.}\ }\textbf
  {\bibinfo {volume} {117}},\ \bibinfo {pages} {016602} (\bibinfo {year}
  {2016}{\natexlab{a}})}\BibitemShut {NoStop}%
\bibitem [{\citenamefont {Ren}\ \emph {et~al.}(2010)\citenamefont {Ren},
  \citenamefont {Taskin}, \citenamefont {Sasaki}, \citenamefont {Segawa},\ and\
  \citenamefont {Ando}}]{Berry.phase2}%
  \BibitemOpen
  \bibfield  {author} {\bibinfo {author} {\bibfnamefont {Z.}~\bibnamefont
  {Ren}}, \bibinfo {author} {\bibfnamefont {A.~A.}\ \bibnamefont {Taskin}},
  \bibinfo {author} {\bibfnamefont {S.}~\bibnamefont {Sasaki}}, \bibinfo
  {author} {\bibfnamefont {K.}~\bibnamefont {Segawa}}, \ and\ \bibinfo {author}
  {\bibfnamefont {Y.}~\bibnamefont {Ando}},\ }\href {\doibase
  10.1103/PhysRevB.82.241306} {\bibfield  {journal} {\bibinfo  {journal} {Phys.
  Rev. B}\ }\textbf {\bibinfo {volume} {82}},\ \bibinfo {pages} {241306}
  (\bibinfo {year} {2010})}\BibitemShut {NoStop}%
\bibitem [{\citenamefont {Pariari}\ \emph {et~al.}(2015)\citenamefont
  {Pariari}, \citenamefont {Dutta},\ and\ \citenamefont
  {Mandal}}]{Berry.phase3}%
  \BibitemOpen
  \bibfield  {author} {\bibinfo {author} {\bibfnamefont {A.}~\bibnamefont
  {Pariari}}, \bibinfo {author} {\bibfnamefont {P.}~\bibnamefont {Dutta}}, \
  and\ \bibinfo {author} {\bibfnamefont {P.}~\bibnamefont {Mandal}},\ }\href
  {\doibase 10.1103/PhysRevB.91.155139} {\bibfield  {journal} {\bibinfo
  {journal} {Phys. Rev. B}\ }\textbf {\bibinfo {volume} {91}},\ \bibinfo
  {pages} {155139} (\bibinfo {year} {2015})}\BibitemShut {NoStop}%
\bibitem [{\citenamefont {Sergelius}\ \emph {et~al.}(2016)\citenamefont
  {Sergelius}, \citenamefont {Gooth}, \citenamefont {B{\"a}{\ss}ler},
  \citenamefont {Zierold}, \citenamefont {Wiegand}, \citenamefont {Niemann},
  \citenamefont {Reith}, \citenamefont {Shekhar}, \citenamefont {Felser},
  \citenamefont {Yan} \emph {et~al.}}]{Berry.phase4}%
  \BibitemOpen
  \bibfield  {author} {\bibinfo {author} {\bibfnamefont {P.}~\bibnamefont
  {Sergelius}}, \bibinfo {author} {\bibfnamefont {J.}~\bibnamefont {Gooth}},
  \bibinfo {author} {\bibfnamefont {S.}~\bibnamefont {B{\"a}{\ss}ler}},
  \bibinfo {author} {\bibfnamefont {R.}~\bibnamefont {Zierold}}, \bibinfo
  {author} {\bibfnamefont {C.}~\bibnamefont {Wiegand}}, \bibinfo {author}
  {\bibfnamefont {A.}~\bibnamefont {Niemann}}, \bibinfo {author} {\bibfnamefont
  {H.}~\bibnamefont {Reith}}, \bibinfo {author} {\bibfnamefont
  {C.}~\bibnamefont {Shekhar}}, \bibinfo {author} {\bibfnamefont
  {C.}~\bibnamefont {Felser}}, \bibinfo {author} {\bibfnamefont
  {B.}~\bibnamefont {Yan}},  \emph {et~al.},\ }\href
  {https://www.nature.com/articles/srep33859?WT.feed_name=subjects_topological-insulators}
  {\bibfield  {journal} {\bibinfo  {journal} {Scientific reports}\ }\textbf
  {\bibinfo {volume} {6}},\ \bibinfo {pages} {33859} (\bibinfo {year}
  {2016})}\BibitemShut {NoStop}%
\bibitem [{\citenamefont {Lifshitz}\ and\ \citenamefont
  {Kosevich}(1956)}]{LK1}%
  \BibitemOpen
  \bibfield  {author} {\bibinfo {author} {\bibfnamefont {I.~M.}\ \bibnamefont
  {Lifshitz}}\ and\ \bibinfo {author} {\bibfnamefont {A.~M.}\ \bibnamefont
  {Kosevich}},\ }\href
  {http://www.ujp.bitp.kiev.ua/files/journals/53/si/53SI25p.pdf} {\bibfield
  {journal} {\bibinfo  {journal} {Sov. Phys. JETP}\ }\textbf {\bibinfo {volume}
  {2}},\ \bibinfo {pages} {636} (\bibinfo {year} {1956})}\BibitemShut {NoStop}%
\bibitem [{\citenamefont {Shoenberg}(2009)}]{shoenburg}%
  \BibitemOpen
  \bibfield  {author} {\bibinfo {author} {\bibfnamefont {D.}~\bibnamefont
  {Shoenberg}},\ }\href@noop {} {\bibfield  {journal} {\bibinfo  {journal}
  {Cambridge University Press, Cambridge, UK}\ } (\bibinfo {year}
  {2009})}\BibitemShut {NoStop}%
\bibitem [{\citenamefont {Hu}\ \emph {et~al.}(2016{\natexlab{b}})\citenamefont
  {Hu}, \citenamefont {Tang}, \citenamefont {Liu}, \citenamefont {Liu},
  \citenamefont {Zhu}, \citenamefont {Graf}, \citenamefont {Myhro},
  \citenamefont {Tran}, \citenamefont {Lau}, \citenamefont {Wei},\ and\
  \citenamefont {Mao}}]{LK3}%
  \BibitemOpen
  \bibfield  {author} {\bibinfo {author} {\bibfnamefont {J.}~\bibnamefont
  {Hu}}, \bibinfo {author} {\bibfnamefont {Z.}~\bibnamefont {Tang}}, \bibinfo
  {author} {\bibfnamefont {J.}~\bibnamefont {Liu}}, \bibinfo {author}
  {\bibfnamefont {X.}~\bibnamefont {Liu}}, \bibinfo {author} {\bibfnamefont
  {Y.}~\bibnamefont {Zhu}}, \bibinfo {author} {\bibfnamefont {D.}~\bibnamefont
  {Graf}}, \bibinfo {author} {\bibfnamefont {K.}~\bibnamefont {Myhro}},
  \bibinfo {author} {\bibfnamefont {S.}~\bibnamefont {Tran}}, \bibinfo {author}
  {\bibfnamefont {C.~N.}\ \bibnamefont {Lau}}, \bibinfo {author} {\bibfnamefont
  {J.}~\bibnamefont {Wei}}, \ and\ \bibinfo {author} {\bibfnamefont
  {Z.}~\bibnamefont {Mao}},\ }\href {\doibase 10.1103/PhysRevLett.117.016602}
  {\bibfield  {journal} {\bibinfo  {journal} {Phys. Rev. Lett.}\ }\textbf
  {\bibinfo {volume} {117}},\ \bibinfo {pages} {016602} (\bibinfo {year}
  {2016}{\natexlab{b}})}\BibitemShut {NoStop}%
\bibitem [{\citenamefont {Xiong}\ \emph {et~al.}(2012)\citenamefont {Xiong},
  \citenamefont {Luo}, \citenamefont {Khoo}, \citenamefont {Jia}, \citenamefont
  {Cava},\ and\ \citenamefont {Ong}}]{landau.fan1}%
  \BibitemOpen
  \bibfield  {author} {\bibinfo {author} {\bibfnamefont {J.}~\bibnamefont
  {Xiong}}, \bibinfo {author} {\bibfnamefont {Y.}~\bibnamefont {Luo}}, \bibinfo
  {author} {\bibfnamefont {Y.}~\bibnamefont {Khoo}}, \bibinfo {author}
  {\bibfnamefont {S.}~\bibnamefont {Jia}}, \bibinfo {author} {\bibfnamefont
  {R.~J.}\ \bibnamefont {Cava}}, \ and\ \bibinfo {author} {\bibfnamefont
  {N.~P.}\ \bibnamefont {Ong}},\ }\href {\doibase 10.1103/PhysRevB.86.045314}
  {\bibfield  {journal} {\bibinfo  {journal} {Phys. Rev. B}\ }\textbf {\bibinfo
  {volume} {86}},\ \bibinfo {pages} {045314} (\bibinfo {year}
  {2012})}\BibitemShut {NoStop}%
\bibitem [{\citenamefont {Hasan}\ and\ \citenamefont
  {Kane}(2010{\natexlab{b}})}]{Rev.TI.Hasan}%
  \BibitemOpen
  \bibfield  {author} {\bibinfo {author} {\bibfnamefont {M.~Z.}\ \bibnamefont
  {Hasan}}\ and\ \bibinfo {author} {\bibfnamefont {C.~L.}\ \bibnamefont
  {Kane}},\ }\href {\doibase 10.1103/RevModPhys.82.3045} {\bibfield  {journal}
  {\bibinfo  {journal} {Rev. Mod. Phys.}\ }\textbf {\bibinfo {volume} {82}},\
  \bibinfo {pages} {3045} (\bibinfo {year} {2010}{\natexlab{b}})}\BibitemShut
  {NoStop}%
\bibitem [{\citenamefont {Lawson}\ \emph {et~al.}(2012)\citenamefont {Lawson},
  \citenamefont {Hor},\ and\ \citenamefont {Li}}]{QO.in.CuBiSe}%
  \BibitemOpen
  \bibfield  {author} {\bibinfo {author} {\bibfnamefont {B.~J.}\ \bibnamefont
  {Lawson}}, \bibinfo {author} {\bibfnamefont {Y.~S.}\ \bibnamefont {Hor}}, \
  and\ \bibinfo {author} {\bibfnamefont {L.}~\bibnamefont {Li}},\ }\href
  {\doibase 10.1103/PhysRevLett.109.226406} {\bibfield  {journal} {\bibinfo
  {journal} {Phys. Rev. Lett.}\ }\textbf {\bibinfo {volume} {109}},\ \bibinfo
  {pages} {226406} (\bibinfo {year} {2012})}\BibitemShut {NoStop}%
\end{thebibliography}%

\end{document}